\documentclass{article}

\usepackage{arxiv}

\usepackage[utf8]{inputenc} 
\usepackage[T1]{fontenc}    
\usepackage{hyperref}       
\usepackage{url}            
\usepackage{booktabs}       
\usepackage{amsfonts}       
\usepackage{nicefrac}       
\usepackage{microtype}      
\usepackage{lipsum}
\usepackage{graphicx}
\usepackage{float}
\graphicspath{ {./figures/} }

\title{An Adaptable and Agnostic Flow Scheduling Approach for Data Center Networks}

\author{
 Sergio Armando Guti\'errez \\
  Facultad de Ingenier\'ias\\
  Universidad Aut\'onoma Latinoamericana\\
  Carrera 55 N° 49-51, Medell\'in, Colombia \\
  \texttt{sergio.gutierrezbe@unaula.edu.co} \\
   \And
 Juan Felipe Botero \\
  Facultad de Ingenier\'ia\\
  Universidad de Antioquia\\
  Carrera 75 N° 65-87, Medell\'in, Colombia  \\
  \texttt{juanf.botero@udea.edu.co} \\
  \And
 John Willian Branch \\
  Facultad de Minas\\
  Universidad Nacional de Colombia\\
  Carrera 80 N° 65-223, Medell\'in, Colombia \\
  \texttt{jwbranch@udea.edu.co} \\
}
\def\makeheadbox{{%
\hbox to0pt{\vbox{\baselineskip=10dd\hrule\hbox
to\hsize{\vrule\kern3pt\vbox{\kern3pt
\hbox{\bfseries [Insert journal name here]}
\hbox{This is a pre-peer review, pre-print version of this article.}
\kern3pt}\hfil\kern3pt\vrule}\hrule}%
\hss}}}

\begin{document}
\maketitle
\begin{abstract}
Cloud applications have reshaped the model of services and infrastructure of the Internet. Search engines, social networks, content delivery and retail and e-commerce sites belong to this group of applications. An important element in the architecture of data centers where these applications run is the communication infrastructure, commonly known as data center networks (DCNs). A critical challenge DCNs have to address is the processing of the traffic of cloud applications, which due to its properties is essentially different to the traffic of other Internet applications. In order to improve the responsiveness and throughput of applications, DCNs should be able to prioritize short flows (a few KB) over long flows (several MB). However, given the time and space variations the traffic presents, the information about flow sizes is not available in advance in order to plan the flow scheduling. There has been a wealth of solutions developed in this space, and prior work includes flow scheduling mechanisms optimizing for a specific workload but fall short when workloads are not known in advance, or comprise a collection of applications changing dynamically. In this paper, we present an adaptable mechanism called Adaptable Workload-Agnostic Flow Scheduling (AWAFS). It is an adaptable approach that can adjust in an agnostic way the scheduling configuration of DCN forwarding devices. This agnostic adjustment contributes to reduce the Flow Completion Time (FCT) of those short flows, representing around 85\% of the traffic handled by cloud applications. AWAFS operates by observing the traffic and detecting statistical properties that provide a hint to adapt the scheduling parameters. Our evaluation results based on simulation show that AWAFS can reduce the average FCT of short flows between 16.9\% and 45.2\% when compared to the best existing agnostic non-adaptable solution, without inducing starvation on long flows. Indeed, it can provide improvements as high as 39\% for long flows. Additionally, AWAFS can improve the FCT for short flows in scenarios with high heterogeneity in the traffic present in the network, with a reduction up to 5\% for the average FCT and 15\% for the tail FCT. 
\end{abstract}


\section{Introduction}
Many cloud applications running on Data Center Networks (DCNs) have very stringent latency requirements. The satisfaction of these requirements impacts the user perception and therefore the revenue obtained by the owners of such applications
\cite{chen_2018,sriraman_2018,xia_survey_2017,rojas-cessa_schemes_2015, brutlag_speed_2009}. From a traffic engineering perspective, the traffic associated with these applications consists of a mix of mice flows (those transporting a few Kilobytes) and elephant flows (those transporting several Megabytes or Gigabytes). Responsiveness, as perceived by users, is associated to mice flows whereas quality and completeness of the output provided by applications is associated to elephant flows \cite{amezquita_2019,joy_improving_2015}.
There are three important performance goals related to cloud applications, aiming at achieving the required levels of responsiveness and output quality. These are: deadline accomplishment for time-constrained flows, minimization of Flow Completion Time (FCT) for mice flows, and high throughput without starvation for elephant flows \cite{handley_2017,ruan_2017,chen_scheduling_2016}.

There are two main approaches to address the above goals: Queue management and flow scheduling \cite{liu_2018,hafeez_2017,wang_2016}. Queue management tries to reduce the delays due to occupation in switch buffers (queuing delays) affecting specially mice flows. Different approaches propose ideas such as reservation of buffer space to face traffic bursts \cite{Lu_2017,alizadeh_less_2012}, controlling rate transmission at end hosts \cite{sharma2018approximating,mittal_2015,munir_minimizing_2013,alizadeh_data_2010} or even bypassing the switch queues under specific circumstances \cite{grosvenor_queues_2015}.

On the other hand, flow scheduling aims at controlling how the switches should perform the packet scheduling in order to achieve the mentioned performance goals (deadline accomplishment, FCT minimization for mice flows and high throughput for elephant flows) \cite{rojas-cessa_schemes_2015}. From the scheduling perspective, the simultaneous achievement of those goals introduces three challenges. First, for many data center applications, it is difficult (or even impossible) to deliver information to the transport layer in order to enable close-to-optimal flow scheduling. This limitation arises due to the non-trivial modifications required in the Operating Systems of hosts and the applications themselves \cite{alizadeh_pfabric:_2013}. Second, in many cases, it is not possible to have a-priori information needed to plan the flow scheduling for a given application or even a mix of applications \cite{dukic_2019, bai_pias:_2017,chowdhury_efficient_2015}. Third, it is not possible to have a scheduling mechanism tailored for each application running on a data center, capable to deal with time and space variations that the traffic might exhibit \cite{chen_2018,chen_scheduling_2016}.

Despite related work presents different proposals that try either approximating optimal scheduling algorithms minimizing the average FCT for the applications or maximizing deadline meeting \cite{bai_pias:_2017, chen_scheduling_2016,munir_friends_2014,alizadeh_pfabric:_2013,hong_finishing_2012}, these proposals fail in addressing the previously mentioned challenges \cite{oljira_2018}. In particular, they assume that properties such as the flow size distribution for the workloads present on the network can be known in advance \cite{bai_pias:_2017,chen_scheduling_2016}, which in practice is difficult to achieve with reasonable levels of accuracy \cite{dukic_2019}.

In this paper, we propose a mechanism that satisfies two performance goals of cloud applications: minimization of the FCT for mice flows and high throughput without starvation for elephant flows. The mechanism is called Adaptable Workload-Agnostic Flow Scheduling, or AWAFS. In AWAFS, hosts piggyback information in packets. This information allows forwarding devices to dynamically adapt the flow scheduling process to reduce the FCT of mice flows. This adaption does not require prior information about traffic properties of the workloads present on the network. The workload agnosticism of AWAFS regarding flow sizes contributes to adequately handle the spatial and time variations usually exhibited by the traffic associated to cloud applications \cite{roy_inside_2015,benson_network_2010,alizadeh_data_2010}.

The remaining of this paper is organized as follows:
Section \ref{related} presents a review of related work in the problem of minimization of Flow Completion Times in DCN. Section \ref{awafs} introduces the design of our proposal for workload-agnostic flow scheduling, that we have called AWAFS. Section \ref{evaluation} shows the results of the evaluation performed to assess our proposal. In Section \ref{discussion} we present a discussion of the insights obtained with the development of AWAFS. Finally, in Section \ref{Conclusions} we state the conclusions and future work.

\section{Related Work}
\label{related}
Previous work on minimization of FCT in DCN can be divided in two main categories: Information-aware  and Information-agnostic solutions \cite{rojas-cessa_schemes_2015,noormohammadpour_datacenter_2017}. 

Information-aware mechanisms leverage information received in advance from applications. This information can represent deadlines to be satisfied or the actual size of flows. This allows to implement in detail the flow scheduling \cite{bai_pias:_2017,rojas-cessa_schemes_2015}. In this category we can mention PDQ \cite{hong_finishing_2012} which achieves quick flow completion and deadline satisfaction, pFabric \cite{alizadeh_pfabric:_2013} which decouples flow scheduling and rate control and PASE \cite{munir_friends_2014}, which provides a combined solution integrating elements such as self-adjusting points, arbitration and in-network prioritization.

On the other hand, Information-agnostic approaches do not require to have in advance explicit information about either flow sizes or deadlines. This is addressed either by controlling queue lengths as in DCTCP \cite{alizadeh_data_2010}, HULL \cite{alizadeh_less_2012}, QJUMP \cite{grosvenor_queues_2015}, or by adjusting the scheduling mechanisms within the switches as in NDP \cite{handley_2017,zilberman2020artifact}, PIAS \cite{bai_pias:_2017,bai_information-agnostic_2015} and KARUNA \cite{chen_scheduling_2016}, which integrates information-aware scheduling to provide deadline-satisfaction if deadline information is available and information-agnostic scheduling for remaining flows.

Information-aware approaches rely on either having explicit information about flow sizes \cite{alizadeh_pfabric:_2013,hong_finishing_2012} or including additional elements for arbitration and control of resource allocation \cite{munir_friends_2014}. There are two main reasons that hinder the practical adoption of these approaches. First, despite some cloud applications might provide in advance the information of flow sizes \cite{peng_hadoopwatch:_2014}, modifications at data center infrastructure to consider this information in the packet scheduling are prohibitive \cite{alipio2019tcp,noormohammadpour_datacenter_2017, zhang2016minimizing}. These modifications include patching application code, altering kernel code to pass flow sizes to transport layer protocols or even changing the behavior of the switches forming the data center fabric, which are nowadays mostly commodity hardware \cite{bai_pias:_2017}. Second, the inclusion of additional elements in the network control plane \cite{munir2016pase} increases the complexity and might introduce additional points of failure in the infrastructure.

In contrast, PIAS \cite{bai_pias:_2017} and KARUNA \cite{chen_scheduling_2016} do not use prior information about the size of a flow in order to schedule it. Despite this agnostic operation, they still do need prior information about the Cumulative Distribution Function (CDF) of the flow sizes of the workloads present in the network and their estimated traffic load in order to define the parameters (i.e. the demotion thresholds) for their MLFQ-based scheduling component.

Due to the time and space variations exhibited by the traffic present in DCN \cite{dukic_2019,amezquita_2019,noormohammadpour_datacenter_2017, roy_inside_2015}, a challenging task associated to MLFQ-based scheduling is the derivation of a set of demotion thresholds that minimizes the average and tail FCT. In order to perform this task, PIAS calculates these thresholds based on traffic information consisting in the CDF of the flow sizes of the workload that is expected to be present in the network. When the thresholds are derived, they are distributed and deployed at end hosts. These thresholds should match priority levels at switches. Then, end hosts use these thresholds to perform packet tagging according to the priority values associated to the queues at switches \cite{bai_pias:_2017}. 

There are two main limitations on the approach followed by PIAS (and the component of KARUNA dealing with non deadline-constrained traffic):

\begin{itemize}

\item A set of thresholds that might minimize the average FCT in a segment of the network might not be adequate at other segments.

\item The demotion thresholds are derived a-priori (e.g. by a controller) from the CDF of a given expected workload, and they need to be manually deployed at end hosts. Hence, they are not autonomously adjusted and deployed upon the arrival of a different workload.  
\end{itemize}

\begin{figure}[htp]%
    \centering
    {\includegraphics[width=0.65\linewidth]{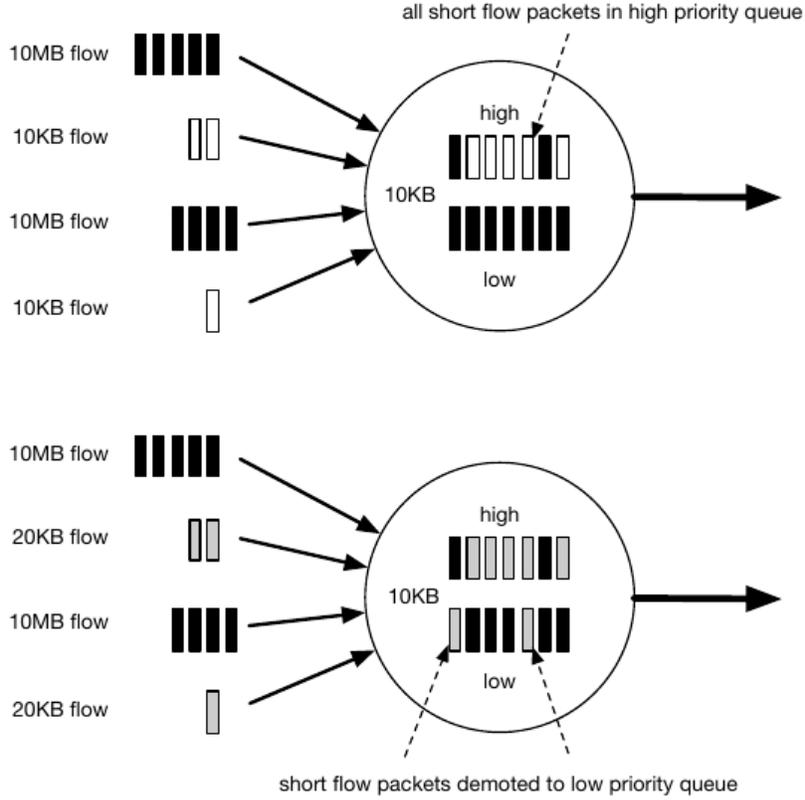} }%
    \caption{The Threshold Mismatch problem}
    \qquad
    \label{fig:mismatch}%
\end{figure}

\textbf{Motivating Example:} The limitations previously described ultimately lead to what we call the Threshold Mismatch problem. This is a problem which might hamper the goal of minimizing the FCT, hurting specially short flows \cite{chen_2018}. The Threshold Mismatch problem can be explained with the example illustrated on Figure \ref{fig:mismatch}. 

Assume a simplified MLFQ with 2 priority queues and, therefore, a single demotion threshold. Suppose that, for the sake of simplicity, there are only two flow sizes: 10KB and 10MB, and that for a given workload, 90\% of its flows are 10KB and the remaining 10\% are 10MB. In this situation, setting the demotion threshold to 10KB will be the optimal configuration to reduce the FCT of the short flows since it will keep them in the highest priority queue until their completion. Thus, short flows result prioritized over the long flows, which will be ultimately enqueued into the low priority queue. Now, suppose there is a shift in the workload, and the size of small flows becomes 20KB instead of 10KB. With the demotion threshold set to 10KB, some of the packets of the short flows will be demoted to the low priority queue. In this situation, those packets will be mixed with the packets of the 10MB flows. This will cause that the latency for the short flows increases due to the competition of their packets with those of the long flows.

The aforementioned situation might happen both either in a later time (time variation) or at a different switch in the network (space variation).

\textbf{Envisaged Deployment:} The Threshold Mismatch problem shows the need of a different, self-adaptive scheme, that runs on every forwarding device of the network. This mechanism should be able to observe the traffic in order to adapt the MLFQ thresholds accordingly.

\textbf{Contribution:} Our main contribution in this paper is providing a MLFQ-based scheduling system with self-adaptability capabilities. AWAFS can adapt its configuration in order to react to time and space variations in the traffic, without requiring a-priori information about the traffic properties. 

\section{AWAFS: ADAPTABLE WORKLOAD-AGNOSTIC FLOW SCHEDULING}\label{awafs}

\textbf{Main Concept:} As previously mentioned, AWAFS is a mechanism conceived to run at switches in order to adjust the demotion thresholds of the scheduling mechanism. Similar to other related solutions on Flow Scheduling \cite{kundel2018p4,bai_pias:_2017,chen_scheduling_2016,alizadeh_less_2012}, AWAFS is based on the use of Multi Level Feedback Queue (MLFQ). In MLFQ, flows are dynamically demoted from higher priority queues towards low priority queues according to the amount of bytes sent. The more bytes a flow transmits, the lower priority queue its packets are queued. By using MLFQ, AWAFS aims at emulating the Shortest Job First heuristic, which is known to be the optimal scheduling discipline to minimize FCT in a single link. This approach has been also followed by previous related work \cite{bai_pias:_2017}. A key difference between AWAFS and previous related work is the capability to dynamically adapt the demotion thresholds of the MLFQ scheduler. In order to enable the adaptability of these thresholds, the switch stores information about recent flows completed. More precisely, it keeps a finite list of tuples, each one containing the final size and the timestamp of the last packet in a flow. Periodically, the switch flushes the list so that tuples with timestamps older than some value are excluded, implementing an \textbf{observation window}.  


\subsection{Detailed Mechanism}
\label{detailed}

\textbf{Packet marking at end hosts:} AWAFS leverages information provided by end hosts to perform the adjustment of its scheduling configuration at the forwarding devices. In order to perform the calculations required to adjust the demotion thresholds, end hosts mark packets at the end of each flow to let switches learn about its final size. This notification can be performed either by defining a custom header like in similar work \cite{bai_pias:_2017,alizadeh_pfabric:_2013,hong_finishing_2012,wilson_better_2011} or by leveraging features of programmable data planes \cite{castanheira2019flowstalker,bosshart_p4:_2014,sivaraman_dc.p4:_2015} to store some state for the flows and later inserting such header. 


\textbf{Functionalities at forwarding devices:} The proposed design relies on the following features in forwarding devices:

\begin{itemize}
    \item Keep track of the amount of bytes forwarded for each specific flow (e.g. by using counters).
    \item Detect the event of flow completion (i.e. the end of a flow). 
    \item For each switch port, keep a data structure to store 2-tuples containing i) the final size of a completed flow and ii) the timestamp of the flow completion event. 
    \item Calculate a set of percentiles for a given set of values.  
\end{itemize}


\textbf{Queue Selection:} AWAFS is based on a MLFQ scheduler for each switch port. This scheduler consists of $k$ unbounded priority queues $Q_{k}$, $k > 1$ with an associated set of demotion thresholds $Thr_{i}$, $ 0 \leq i \leq k-1$. Let be $Q_{1}$ the highest priority queue, and $Q_{k}$ the lowest priority queue (i.e. the queues represent decreasing processing priority). Let $Thr_{i}$, $0 \leq i \leq k-1$ be the demotion threshold between $Q_{i}$ and $Q_{i+1}$. The set of values $\{Thr_{i}\}$ represents byte counts. Whenever a flow starts, its packets are enqueued at $Q_{1}$. The byte flow count is increased with each packet, and when an arriving packet makes the counter exceed $Thr_{0}$, then this packet is enqueued at $Q_{2}$. In general, when the flow has sent more than $Thr_{i}$ bytes, the packets are enqueued at $Q_{i+1}$. If a flow informs that it has sent more than $Thr_{k-1}$ bytes, then its packets are marked to be enqueued into the $Q_{k}$ queue (i.e. the lowest priority queue). 

\textbf{Periodic Adaption:} The traffic incoming at a Forwarding Device will provide information about the workload present in that part of the network. Periodically, each forwarding device uºpdates the demotion thresholds of the MLFQ scheduler of each port. The information contained in the observation window is used for the updating process. Specifically, from the list of sizes of the flows completed during the observation window, a set of percentiles is calculated and used to adjust the demotion thresholds. Low percentiles can define upper bounds for short flows, which should be associated to higher priority levels whereas higher percentiles can define upper bounds for long flows, which should be associated to lower priority levels.

\begin{figure}[htp]%
    \centering
    {\includegraphics[width=0.65\linewidth]{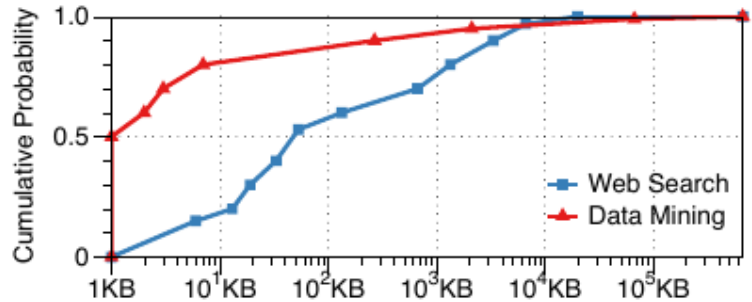} }%
    \caption{Typical data center workloads \cite{bai_pias:_2017}}
    \qquad
    \label{fig:example}%
\end{figure}

\textbf{Supporting example:} Consider the workload Data Mining (red curve) which is shown on Figure \ref{fig:example}. Assume a simple MLFQ system with two queues (therefore, one demotion threshold). For that workload, the plotted Cumulative Probability indicates that approximately 80\% of the flows are shorter than 10KB. Intuitively, by calculating low percentiles (e.g. 10th or 20th percentile) in the list of completed flow sizes, we might have a good hint to set a demotion threshold approximating the upper bound for the short flows in the workload. Hence, the short flows will be prioritized by separating them from the long flows.

\subsubsection{Switch design}
\label{awafs:architecture}
The AWAFS mechanism in a switch is structured in three components: Sensor, Actuator and Scheduler. The Sensor component monitors the traffic to obtain information provided by end hosts and stores it in custom data structures within the switch during a given observation window. The Actuator component leverages the information collected by the Sensor component, and uses it to adjust the demotion thresholds of the Scheduler. The Scheduler component is on charge of performing the flow scheduling. 


\begin{figure}[htp]%
    \centering
    {\includegraphics[width=0.65\linewidth]{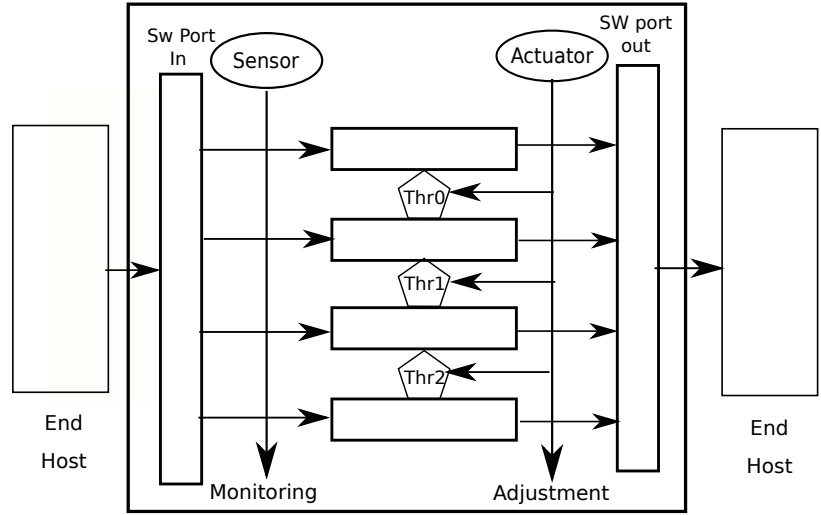} }%
    \caption{Architecture of AWAFS with a configuration of 4 priority queues}
    \qquad
    \label{fig:awafs}%
\end{figure}

\textbf{Sensor:} The Sensor component processes the packets containing the notification of flow completion sent by the end hosts. This information is stored in custom data structures which will be lately used by the Actuator in order to perform the demotion threshold adjustment.

\textbf{Data structures:} In the operation of AWAFS, forwarding devices need to keep a data structure in memory, associated to each port. This data structure is the list of completed flow sizes. It is implemented as a singly-linked list containing 2-tuples $ < ts, size >$. This list is populated upon the event of flow completion is detected (e.g. by receiving a packet marked from the end host). The $ts$ field is the time stamp when the notification of flow completion is received at switch and the $size$ field contains the size of the completed flow. This final size is determined either by explicit notification from the end host, or by performing sequence analysis at the switch. In the next section, we present some results analyzing the possible overhead that this data structure might introduce in the switch operation. 

\textbf{Actuator:} The Actuator components performs the adjustment of the demotion thresholds. This adjustment consists in two main tasks:

\begin{itemize}
    \item First, it prunes the data structure containing the sizes of the completed flows by eliminating those entries out of the scope of the observation window (i.e. older than the observation period). For example, assume that the adjustment process runs every 100ms, and the sliding window is set 500ms. (That is, it considers all the flow sizes completed during the previous 500ms). Whenever the adjustment is run, the forwarding device eliminates from the list the flow size information older than 500ms. By doing this, the actuator implements the window sliding. 
    \item Second, it calculates the set of percentiles that will be used to update the demotion threshold. In order to perform this task, the actuator needs to sort the entries within the scope of the observation window. This sorting can be implemented either by executing a sort algorithm on the list, or by copying the entries onto an implicitly ordered list. After having this ordered list, the calculation of the percentiles can be performed directly. Upon the calculation, the percentiles are used to update the demotion thresholds of the corresponding port.
\end{itemize}

The implementation of the pruning and the sorting to calculate the percentiles depends on the particularities of the specific forwarding device. However, considering for example the context of programmable switches, an approach that could be considered for its implementation is leveraging the General Purpose CPU of the switch. Since the processing of the information involved in the calculation does not directly impact the actual packet forwarding as it is not performed at the data plane (but in the configuration of the demotion thresholds of the MLFQ scheduler), this approach should not introduce important impact in the packet forwarding performance. 

\textbf{Scheduler:} The scheduler is based on conventional MLFQ. In our implementation of MLFQ, packets in different queues are scheduled with strict priority whereas packets in the same queue are scheduled following a FIFO discipline. Queues are selected according to the logic previously described. Hence, short flows tend to complete at higher priority queues while large flows will eventually by demoted to the lowest priority queue. The use of MLFQ introduces a main advantage. Short flows become prioritized over long flows. Moreover, this prioritization is agnostic to the flow size (i.e. the flow size does not 
need to be known in advance). 

\subsection{Discussion} The approach to achieve adaptability of the demotion thresholds for a MLFQ scheduler introduced by AWAFS relies on features that might not be present in commodity switches. It rather leverages functionalities associated to the context of programmable devices. It is a reality that most of the data centers currently deployed are based on commodity switches. However, the standardized configuration of these devices does not allow the implementation of custom data structures or the execution of specific tasks to perform calculations such as the ones previously described used by AWAFS. Recently, the industry and the academia have started to consider the Programmable Data Planes as a landscape enabling the implementation of custom functionalities for packet processing. Moreover, there are proposals considering additional processing tasks to be deployed as services to be consumed by Programmable Devices \cite{xiong2019switches,mustard2019jumpgate}. Programmable Data Planes provide artifacts that allow to define the details of the packet processing process at forwarding devices \cite{he2019flexibility}. 

There are reports confirming the interest of the industry in adopting programmable forwarding devices both in their fabrics \cite{p4org_2018} and at end hosts \cite{firestone2018azure}. Even large carriers and service providers have incorporated Programmable Forwarding Devices to implement critical functions in their traffic processing facilities \cite{att_2017}. In addition to successful deployments at production scenarios, the scientific community is showing an increased interest in the possibilities offered by this concept \cite{soni2020composing, shi2020intflow, sharma2020programmable,kagami2019capest}, despite some restrictions that still need to be addressed, specially regarding hardware resources \cite{sivaraman2017heavy}. Programmable Data Planes are considered a feasible starting point for the development of custom solutions for traditional networking problems \cite{ding2020incrementally,feldmann2019p4}, and it is expected that in the close future many other novel solutions arise, with evaluations and deployments in production environments \cite{gao2020lyra, sedar2018supporting} due to the interest and support from the industry to the evolution of this concept.

Hence, we claim that the ideas involved in the proposal of AWAFS are feasible to be implemented with state-of-the-art programmable forwarding devices. We consider this is not an aspect limiting a hypothetical deployment. On the contrary, the fact of leveraging artifacts of programmable forwarding devices in its design makes AWAFS a solution aligned with concepts currently developed as part of the state-of-the-art in the area of DCN.

\subsection{Summary}
\label{awafs:summary}
In this section, we have described in detail our proposal. Initially, we identify some limitations present on related work. Then we discuss the design of AWAFS and its main components. We also explained the functionalities that would need to be implemented within a programmable forwarding device in order to implement AWAFS. Finally, we present some evidence to support the fact that an implementation of AWAFS in actual hardware is feasible, given the advances reported both in industry and academia regarding the deployment and adoption of Programmable Data Planes as building block of actual solutions in Data Center Networks.


\section{EXPERIMENTS AND RESULTS}\label{evaluation}

In this section, we present the results of assessing AWAFS. We performed extensive experiments based on simulation in order to evaluate the operation of the mechanism we are proposing. This section is organized in six parts. Part \ref{model} describes the simulation model used for the experiments. We define the input factors and metrics considered in the evaluation, the topology used in the simulation scenario, and the workloads used to draw the flow sizes in the experiments. Part \ref{overhead} discusses the assessment of the overhead introduced by the data structures required by AWAFS. Part \ref{convergence} presents the results of evaluating the convergence of the threshold adjustment mechanism. Part \ref{compared} shows the results of comparing AWAFS against its closest related work, PIAS, considering four different typical workloads in data center applications. Part \ref{heter} extends the comparison of AWAFS against PIAS in an environment with heterogeneous traffic. Finally, in Part \ref{summary} we discuss the achieved results.

\subsection{Simulation Environment}
\label{model}
We evaluated AWAFS via NS-2. This tools has been extensively used for the evaluation of previous related work \cite{alizadeh_pfabric:_2013,bai_pias:_2017}. Our simulations were executed on a server with an Intel(R) Core(TM) i7-4790S CPU with 8 cores @ 3.20Ghz and 8GB of RAM running Linux Debian 64bit. For all the experiments, 30 repetitions were executed. All the values present on the results are displayed within a 95\% confidence interval. Depending on the particular scenario, flows were generated either during a given simulated time, or in a given amount. Details are provided in the description of each experiment. Simulated flows arrived according to a Poisson process and their sources and destinations were chosen randomly among the hosts of the topology. The flow sizes a
were drawn from the corresponding reference workload distributions described below.

\subsubsection{Transport Protocol}
Consistent with the methodology presented in most of the papers in the area of flow scheduling, we used DCTCP \cite{alizadeh_data_2010} with Explicit Congestion Notification (ECN) per-port marking approach. See Section IV-A, number 2 of \cite{bai_pias:_2017} for an analysis and justification for this choice. It is important to remark that AWAFS, similarly to closed related works as PIAS and KARUNA does not modify elements of the transport protocol. Hence, it would be compatible with TCP. DCTCP is preferred instead due to its capacity to react to the extent of congestion instead to the mere presence of it \cite{alizadeh_data_2010}.

\subsubsection{Metrics}
We followed the same line of most related papers \cite{bai_pias:_2017,chen_scheduling_2016,hong_finishing_2012} for the classification of flow sizes. We considered three categories: small, medium and large flows, with sizes  up to 100KB, between 100KB and 10MB, and over 10MB respectively.

Following the methodology of related work \cite{alizadeh_pfabric:_2013,bai_pias:_2017,chen_scheduling_2016}, the main metrics assessed were the average and tail Flow Completion Time (FCT) for each class of flow, as described above. We also considered as complementary metrics the overall average FCT as measurement of the overall performance, and the TCP timeouts count. This metric allows to estimate the overhead that AWAFS might be introducing due to the flow prioritization, and specially the possible occurrence of starvation, which might hamper specially the performance of Long Flows.



\subsubsection{\textbf{Topology}}
\label{fig:descriptionTopology}
\begin{figure}[!htp]%
     \centering
     {\includegraphics[width=1\linewidth]{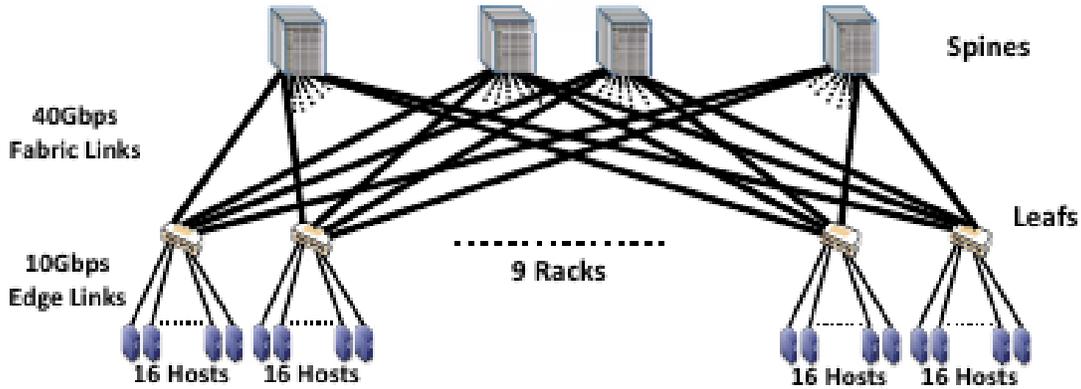} }%
     \caption{Leaf-spine topology used in simulation of the Large Scale scenario}
     \qquad
     \label{fig:topology-largescale}%
 \end{figure}

We used a leaf-spine topology with 9 leaf (ToR) FD connected to 4 spine (Core) FD. In this topology, each leaf switch has 16 10Gbps downlinks (144 hosts) and 4 40Gbps uplinks to the spine, forming a non-oversubscribed network. In our simulation, each FD has 8 queues. Hence, the MLFQ system is based on 8 priority queues and 7 demotion thresholds. The base end-to-end round-trip time across the spine (4 hops) is 85.2 microseconds. We use packet spraying \cite{dixit_impact_2013} for load balancing and disable dupACKs to avoid packet reordering. Figure \ref{fig:topology-largescale} presents a schematic diagram of the simulated topology. 

Aligned with related work \cite{bai_pias:_2017,chen2016features, chen_scheduling_2016,alizadeh_pfabric:_2013} and for the sake of simplicity, we chose this topology. However, since AWAFS is conceived to operate within the FD, it could be incorporated within other topologies such as VL2 or Fat-tree. 

 \subsubsection{Reference Workloads}
 Figures \ref{fig:websearch}, \ref{fig:datamining}, \ref{fig:cache} and \ref{fig:hadoop} present the CDF of the flow sizes for the workloads used for the evaluation. These CDFs are derived and introduced in different previous studies on data center traffic \cite{alizadeh_data_2010,greenberg_vl2:_2009,roy_inside_2015}, and they have been used in the experimental assessment of solutions presented on diverse related work \cite{zilberman2020artifact, mellette2020expanding, cheng2020re,li2019hpcc,chen2019fine, faisal2018workload, bai_pias:_2017,chen_scheduling_2016}


\begin{minipage}{\linewidth}
  \centering
  \begin{minipage}{0.45\linewidth}
      \begin{figure}[H]
          \includegraphics[width=120pt]{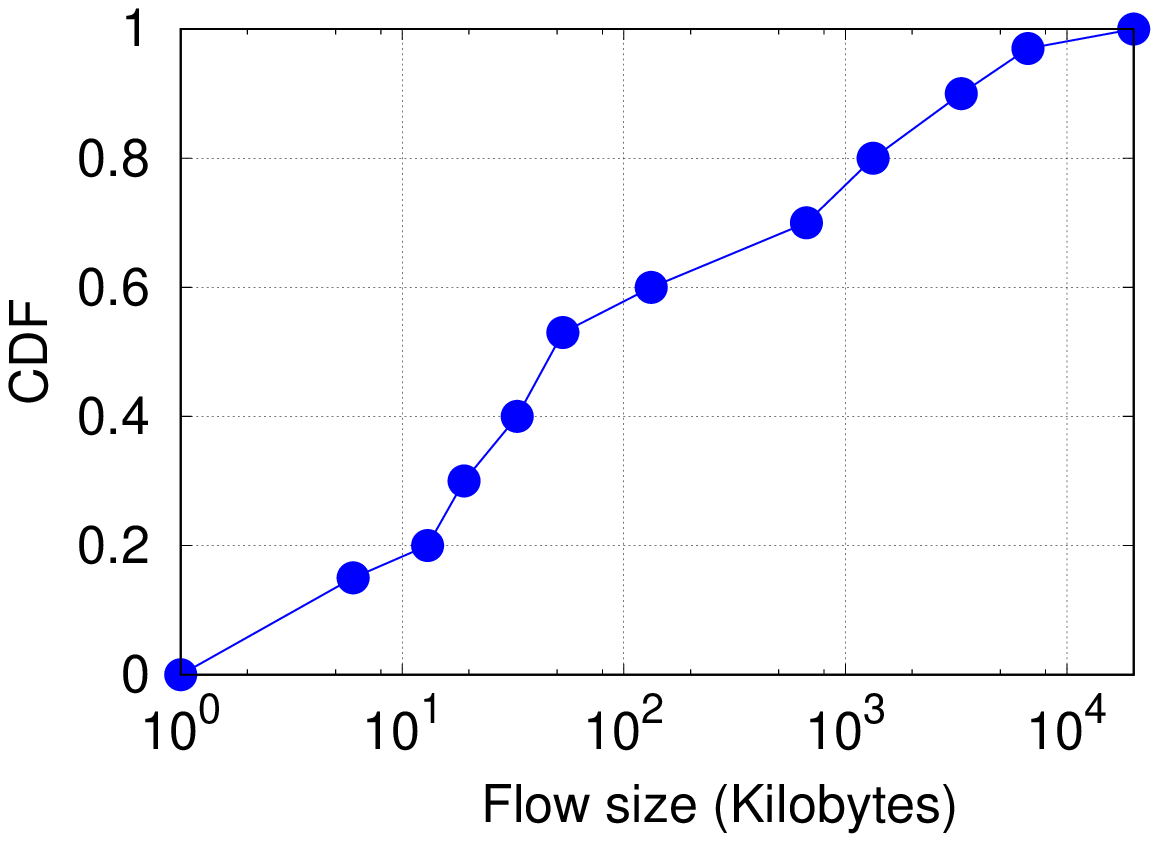}
          \caption{Web Search \cite{alizadeh_data_2010}}
          \label{fig:websearch}
      \end{figure}
  \end{minipage}
  \hspace{0.05\linewidth}
  \begin{minipage}{0.45\linewidth}
      \begin{figure}[H]
          \includegraphics[width=120pt]{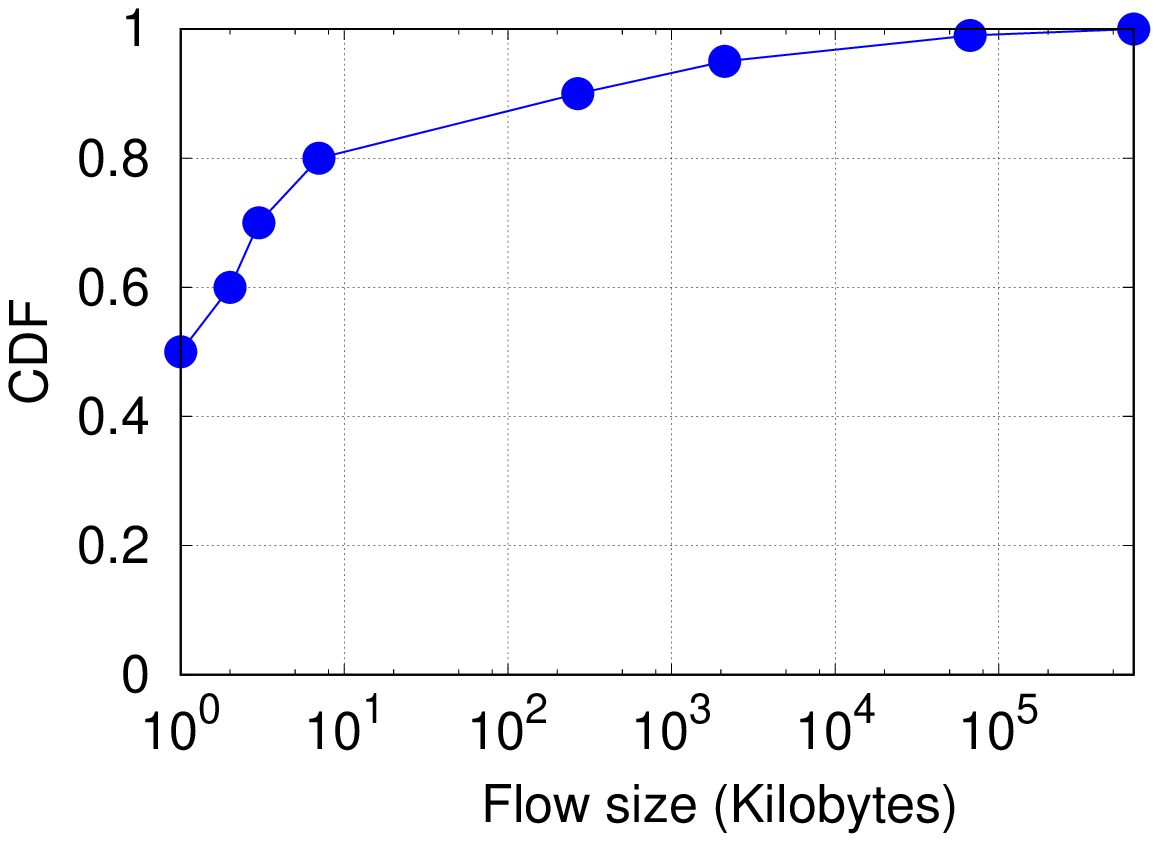}
          \caption{Data Mining \cite{greenberg_vl2:_2009}}
          \label{fig:datamining}
      \end{figure}
      
  \end{minipage}
  \begin{minipage}{0.45\linewidth}
      \begin{figure}[H]
          \includegraphics[width=120pt]{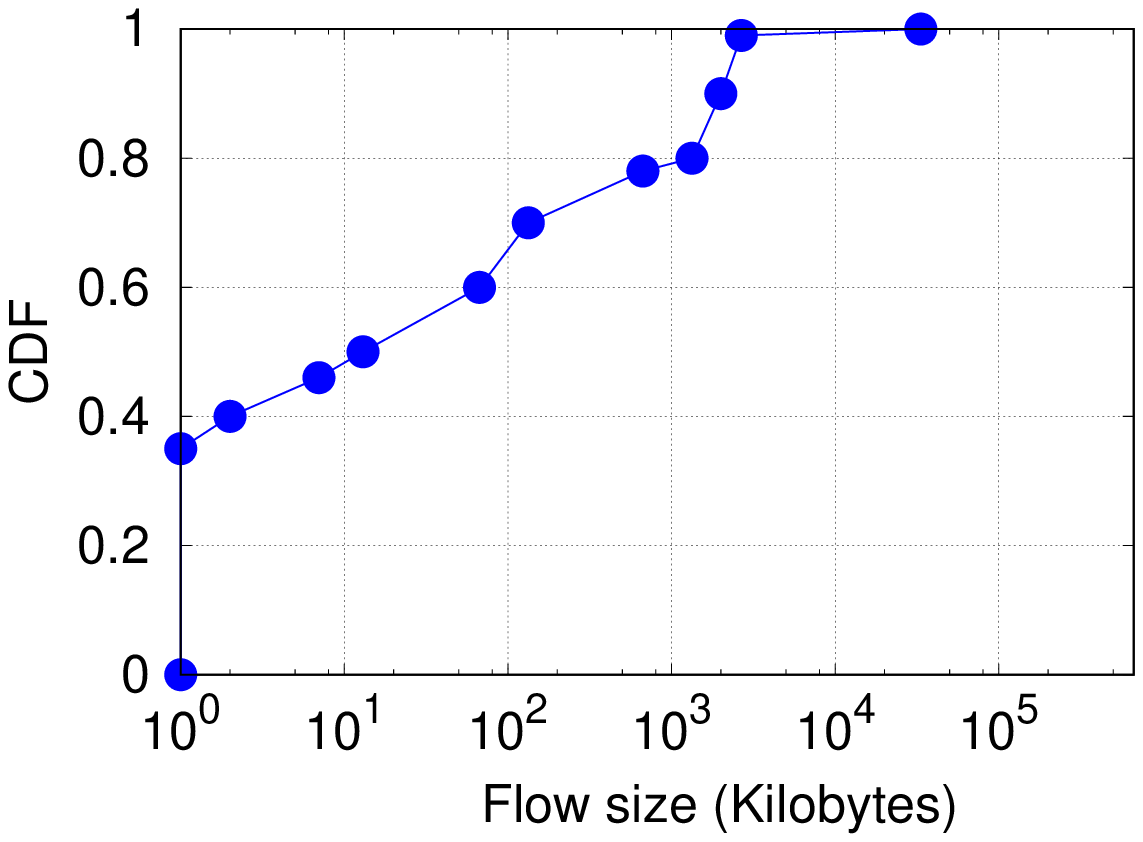}
          \caption{Cache \cite{roy_inside_2015}}
          \label{fig:cache}
      \end{figure}
  \end{minipage}
  \hspace{0.05\linewidth}
  \begin{minipage}{0.45\linewidth}
      \begin{figure}[H]
          \includegraphics[width=120pt]{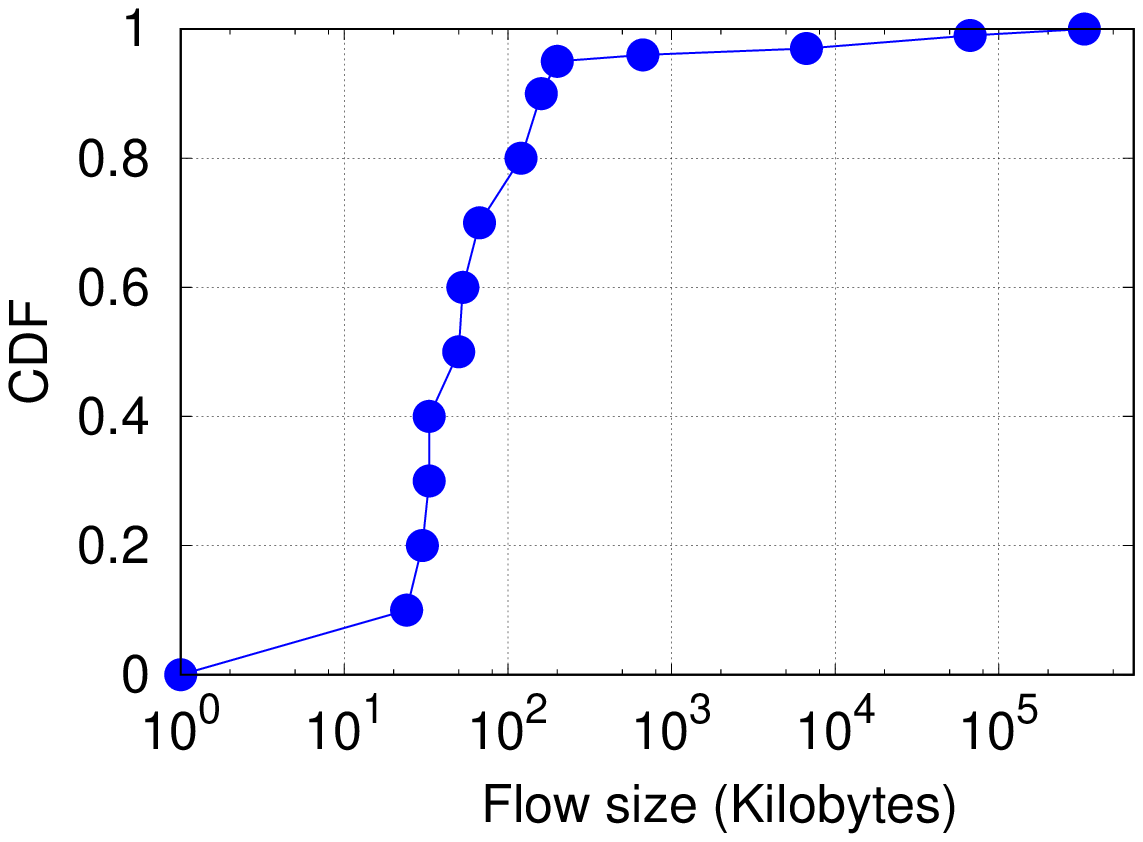}
          \caption{Hadoop \cite{roy_inside_2015}}
          \label{fig:hadoop}
      \end{figure}
      
  \end{minipage}
  
  \end{minipage}

\subsection{Assessment of the Overhead of AWAFS Data Structures}
\label{overhead}
In Section \ref{awafs:architecture}, we described the data structures required for AWAFS operation. The main data structure used is the list for storing the sizes of completed flows. The involvement of this data structure leads to consider its associated overhead. A question to pose is: How much memory overhead would this list introduce in a switch?. In order to answer this question, we conducted a preliminary simulation. We used the simulated leaf-spine topology previously described, and we generated traffic at a load of 90\% of the link capacities during 65s (Simulated time) according to the following description:

\begin{itemize}
\item From T=0 to T=5s, workload Data Mining is executed. This corresponds to the warm-up period.
\item From T=5s to T=35s, workload Data Mining is executed.
\item From T=35s to T=65s, workload Web Search is executed.
\end{itemize}

\begin{figure}[htp]%
    \centering
    
    {\includegraphics[width=0.5\linewidth]{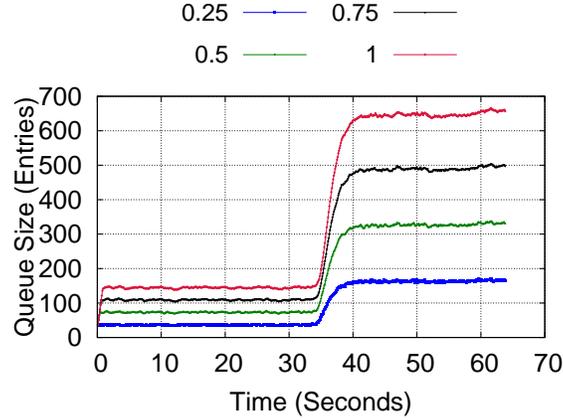} }%
    \caption{Size of the completed flow sizes list}
    \qquad
    \label{fig:qsize}%
\end{figure}

Figure \ref{fig:qsize} presents the average size of the completed flows list observed with four different sizes of the updating window ($W_{update})$: 0.25s, 0.75s, 0.5s and 1s.

With the largest considered window (i.e. 1 second), the size of the list is between 150 and 650 entries. Considering the sizes for the data types required for this list (typically, a 4-byte integer for the size field and a 8-byte double for the timestamp field), and assuming as maximum size the one observed during the experiments, the list occupies around 7.8KB. Then, we can set an upper bound of, for instance, 8KB for the memory used by the list. Hence, AWAFS would require approximately 8K of memory per port in order to store the size of completed flows. 

With respect to the computational complexity of the operations performed on the list, this is an aspect depending on the particular implementation of the associated data structure. In our simulation, we implemented this list as a singly-linked list. It is important to remark that the list is implicitly sorted, since the entries are inserted in time order, whenever a flow is completed (i.e. it is sorted by the time stamp field). Also, the list is defined as singly-linked as it only needs to be traversed in a single direction for the pruning operation. That is, insertions are always performed at the tail of the list and deletions for pruning are performed from the head of the list. Therefore, the list does not need to be traversed to locate where the new entry has to be inserted (i.e. it is inserted at the tail). On the other hand, for the deletion, it is required just to know where the list begins since entries are deleted sequentially until finding the first entry within scope. Considering these two facts, the computational complexity for the insertion and deletion operations in the list is estimated to be $O(1)$. Regarding the complexity of the sorting required for the calculation of the percentiles, it would depend on the particular sorting algorithm used. It might range from $O(nlogn)$ for an algorithm such as HeapSort, or $O(n^2)$ with QuickSort \cite{mehlhorn_data_2013,hopcroft_data_1983}. However, since current standards for Programmable Devices do not support loops \cite{sivaraman_dc.p4:_2015}, this is an operation that would be offloaded onto the General Purpose CPU of the forwarding devices. Therefore, it should not introduce any overhead in packet forwarding.

 \subsection{\textbf{Convergence of the Threshold Adjustment}}
 \label{convergence}
 
 In this experiment, we validated the threshold adjustment mechanism of AWAFS by verifying how it adapted the demotion thresholds upon the occurrence of a workload shift in the network. Using the leaf-spine topology described in subsection \ref{fig:descriptionTopology}, and considering switches with four queues (and therefore three demotion thresholds), we generated traffic at 90\% of the simulated link capacity, according to the following pattern:
 
\begin{itemize}
\item From T=0 to T=5s, workload Data Mining is executed. This corresponds to the warm-up period.
\item From T=5s to T=35s, workload Data Mining is executed.
\item From T=35s to T=65s, workload Web Search is executed.
\end{itemize}

 For the initial values of the demotion thresholds of AWAFS, they were set in such a way that they were optimal for the workload Data Mining, whereas they are suboptimal for the workload Web Search. That is, $Thr_{0} =  7KB$ would cause that approximately 80\% of the flows of the Data Mining workload get completed at the highest priority queue. On the other hand, that threshold would cause that only around 20\% of the flows of the Web Search workload stay at the highest priority queue. That implies an early demotion of the short flows of this workload

The operative parameters of AWAFS were configured as follows:

\begin{itemize}
\item$W\textsubscript{update}$: 1s.

\item$T\textsubscript{schedule}$: 250ms.


\item $\{RefPct_{0}\}: 0.1; RefPct_{i} = RefPct_{i-1}+0.1, 1 \leq i \leq 3$ 

\item $Thr_{0} = 7KB; Thr_{i} = 7 + Thr_{i-1}, 1 \leq i \leq 3$

\end{itemize}

\begin{figure}[htp]%
    \centering
    {\includegraphics[width=0.7\linewidth]{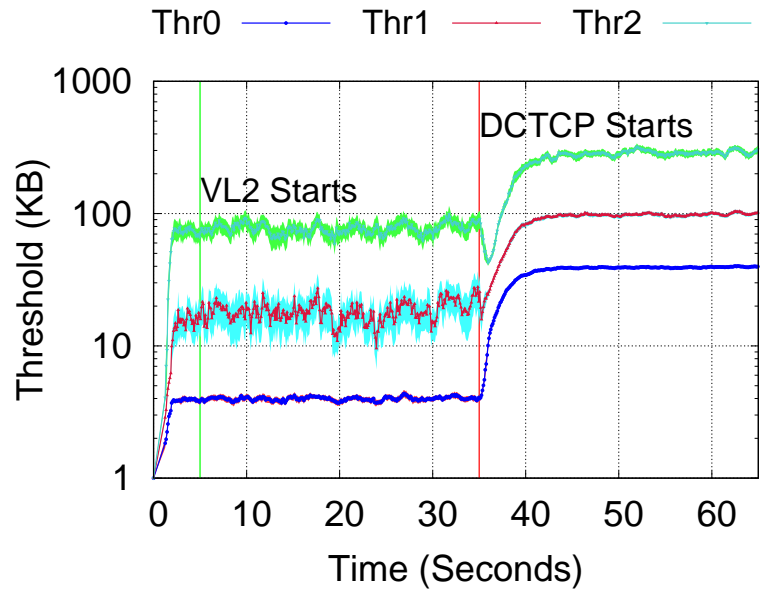} }%
    \caption{Threshold adjustment for 4 priority queues (Logscale)}
    \qquad
    \label{fig:3threshold}%
\end{figure}

Figure \ref{fig:3threshold} shows the adjustment process of the demotion thresholds. Despite the noisy behavior that is observed for Thr2 during the presence of the Data Mining (VL2) workload, it converged to a value closer to 200KB with the Web Search (DCTCP) workload. This noisy behavior can be explained due to the fact that for the Data Mining workload, around 80\% of the flows are shorter than 7KB but the remaining percent of the flows have sizes between 7KB and 600MB. Therefore, the additional priority levels present higher variation of the corresponding percentiles. On the other hand, for the Web Search workload, the range of variations for the corresponding percentile is narrower. Hence, the corresponding thresholds had less noisy behaviors. The colored areas around the curves represent the 95\% interval for the mean value of the given threshold.  

 \subsection{AWAFS vs PIAS with Different Workloads}
\label{compared}

In this part, we compared AWAFS with its closest related work, PIAS \cite{bai_pias:_2017}. In this scenario, we deliberately induced the condition of threshold mismatch. Then, we compared the performance of PIAS with the performance of AWAFS in that situation. In order to induce the threshold mismatch, we proceeded as follows:

\begin{itemize}
\item In the first experiment, traffic was generated using the Web Search workload, and the thresholds generated for the Data Mining workload.
\item In the second experiment, traffic was generated using the Data mining workload, and the thresholds derived for the Web Search workload.
\item In the third and fourth experiments, traffic was generated using the Cache and Hadoop \cite{roy_inside_2015} workloads respectively, and the thresholds generated for the Data Mining workload.
\end{itemize}

The demotion thresholds used in the first and second experiments correspond to those obtained following the procedure described in Section III-C of \cite{bai_pias:_2017} but they are configured in order to induce the threshold mismatch condition. For the third and fourth experiments, we followed the same approach indicated in section V-B of \cite{bai_pias:_2017}. 

For this part of the evaluation, each execution of the experiment consisted in the generation of 100K flows with a given traffic load ranging from 50\% to 90\%. 

\subsubsection{Web Search Workload}

In this section, we present the results achieved by AWAFS compared with PIAS using the demotion thresholds derived for the Data Mining workload. Figures \ref{fig:shortws} and \ref{fig:tailshortws} show the average and tail FCT of short flows respectively. In this experiment, it can be observed that AWAFS outperformed PIAS, with a reduction around 9.6\% of the average FCT and 16.6\% of the tail FCT. At high traffic loads, the improvement introduced by AWAFS was even higher, achieving a reduction of 11.7\% for the average FCT and 15.9\%  for the tail FCT when traffic load was 90\%.

 \begin{minipage}{\linewidth}
  \centering
  \begin{minipage}{0.45\linewidth}
      \begin{figure}[H]
          \includegraphics[width=150pt]{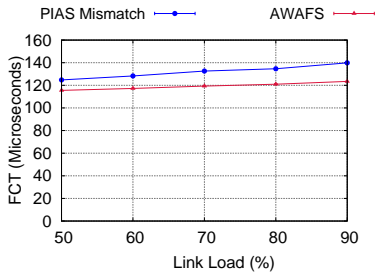}
          \caption{Average FCT of Short Flows - Web Search Workload}
          \label{fig:shortws}
      \end{figure}
  \end{minipage}
  \hspace{0.05\linewidth}
  \begin{minipage}{0.45\linewidth}
      \begin{figure}[H]
          \includegraphics[width=150pt]{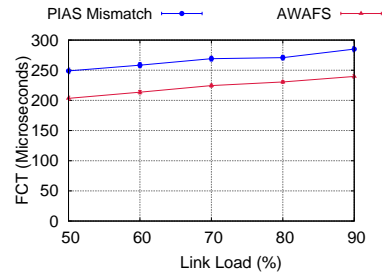}
          \caption{Tail FCT of Short Flows - Web Search Workload}
          \label{fig:tailshortws}
      \end{figure}
      
  \end{minipage}
  \end{minipage}
 \\
 \\

Figures \ref{fig:longws} and \ref{fig:taillongws} present the FCT and tail FCT for long flows. Consistent with the behavior observed for short flows, AWAFS outperformed PIAS and introduced a reduction of up to 20.3\% and 32.7\% for the average and tail FCT of these flows, respectively.

\begin{minipage}{\linewidth}
  \centering
  \begin{minipage}{0.45\linewidth}
      \begin{figure}[H]
          \includegraphics[width=150pt]{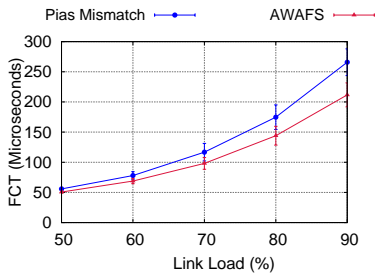}
          \caption{Average FCT of Long Flows - Web Search Workload}
          \label{fig:longws}
      \end{figure}
  \end{minipage}
  \hspace{0.05\linewidth}
  \begin{minipage}{0.45\linewidth}
      \begin{figure}[H]
          \includegraphics[width=150pt]{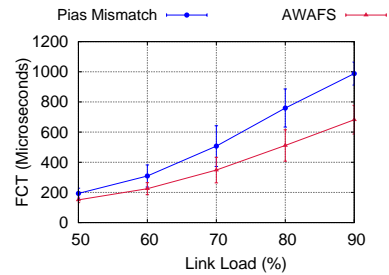}
          \caption{Tail FCT of Long Flows - Web Search Workload}
          \label{fig:taillongws}
      \end{figure}
      
  \end{minipage}
  \end{minipage}
\\
\\
Finally, figures \ref{fig:overallaveragews} and \ref{fig:tcptimeoutsws} present the comparison of the overall average FCT and the count of TCP timeouts, respectively. These results confirm how AWAFS outperformed PIAS by reducing the overall average FCT in 36\% without additional increment of the TCP timeouts. On the contrary, it reduced these events in almost 78\% when compared with the threshold mismatch configuration. This last fact is indication that AWAFS was not creating starvation, which would specially affect long flows. 
\\
\\

 \begin{minipage}{\linewidth}
  \centering
  \begin{minipage}{0.45\linewidth}
      \begin{figure}[H]
          \includegraphics[width=150pt]{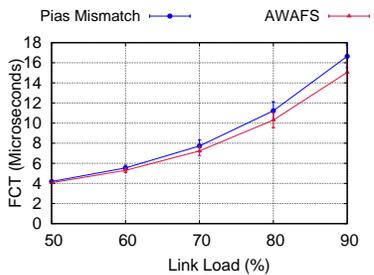}
          \caption{Overall Average FCT - Web Search Workload}
          \label{fig:overallaveragews}
      \end{figure}
  \end{minipage}
  \hspace{0.05\linewidth}
  \begin{minipage}{0.45\linewidth}
      \begin{figure}[H]
          \includegraphics[width=150pt]{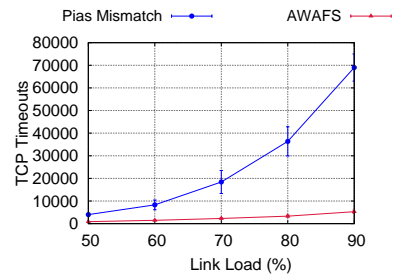}
          \caption{TCP Timeouts Count - Web Search Workload}
          \label{fig:tcptimeoutsws}
      \end{figure}
      
  \end{minipage}
  \end{minipage}

\subsubsection{Data Mining Workload}

In this section, we discuss the results obtained in the experiment with traffic generated with the Data Mining workload with the thresholds generated for the Web Search workload. Considering the properties of this workload, the threshold mismatch does not penalize short flows. It results that the average and tail FCT achieved in the threshold mismatch situation were very close to those obtained with the thresholds generated for the workload, with almost negligible differences. Hence, we observed that the average and tail FCT obtained with AWAFS for the different classes of flows also approximated to the values obtained with the thresholds generated specifically for the workload.

\begin{figure}[htp]%
    \centering
    {\includegraphics[width=0.5\linewidth]{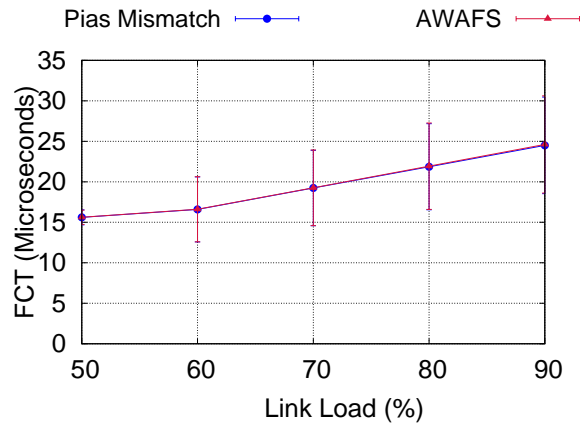} }%
    \caption{Overall Average FCT - Data Mining Workload}
    \qquad
    \label{fig:overallcompareddm}%
\end{figure}

Since the results for this experiment did not present noticeable difference, we only present in Figure \ref{fig:overallcompareddm}, for the sake of comparison, the overall average FCT achieved by AWAFS compared with the achieved by PIAS. It can be observed that PIAS with threshold mismatch is very close to PIAS with the thresholds generated for the workload. AWAFS approximated the average FCT to this value.

 \subsubsection{Hadoop Workload}
 
In this section, we present the results of comparing AWAFS and PIAS when executing the Hadoop workload. In this experiment, we used in PIAS the demotion thresholds derived for the Data Mining workload. 


AWAFS introduced a small improvement for this workload, for the average and tail FCT of the short flows. For the first case, the improvement was between 0.3\% and 0.7\%. For the latter, the improvement was between 0.6\% and 2.3\%. 

Regarding the overall average FCT, it can be observed that the general improvement was small, with a value around 0.4\% for high traffic loads. However, despite the small value of this improvement, AWAFS reduced in almost 95\% the TCP timeouts when compared against PIAS. Figures \ref{fig:overallaveragehadoop} and \ref{fig:tcptimeoutshadoop} present the results of these observations.
  \begin{minipage}{\linewidth}
  \centering
  \begin{minipage}{0.45\linewidth}
      \begin{figure}[H]
          \includegraphics[width=150pt]{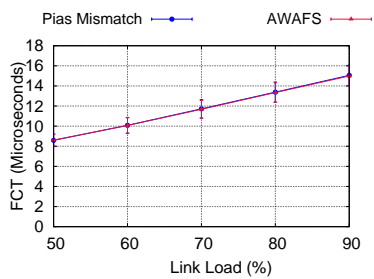}
          \caption{Overall Average FCT - Hadoop Workload}
          \label{fig:overallaveragehadoop}
      \end{figure}
  \end{minipage}
  \hspace{0.05\linewidth}
  \begin{minipage}{0.45\linewidth}
      \begin{figure}[H]
          \includegraphics[width=150pt]{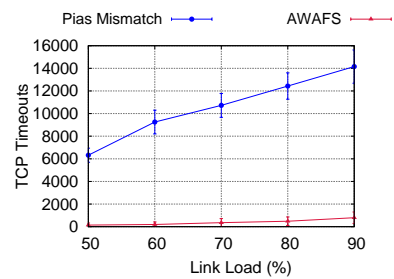}
          \caption{TCP Timeouts Count - Hadoop Workload}
          \label{fig:tcptimeoutshadoop}
      \end{figure}
      
  \end{minipage}
  \end{minipage}



\vspace{0.5cm}
Figures \ref{fig:averagelonghadoop} and \ref{fig:taillonghadoop} present the results of evaluating the FCT for the long flows. AWAFS had a behavior similar to PIAS for these flows. However, it still introduced an improvement close to 1\% at high traffic loads (90\%) for the average FCT and close to 5.7\% for the tail FCT of these flows. 
\\
\\
 \begin{minipage}{\linewidth}
  \centering
  \begin{minipage}{0.45\linewidth}
      \begin{figure}[H]
          \includegraphics[width=150pt]{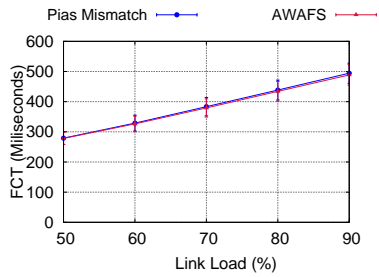}
          \caption{Average FCT for Long Flows - Hadoop Workload}
          \label{fig:averagelonghadoop}
      \end{figure}
  \end{minipage}
  \hspace{0.05\linewidth}
  \begin{minipage}{0.45\linewidth}
      \begin{figure}[H]
          \includegraphics[width=150pt]{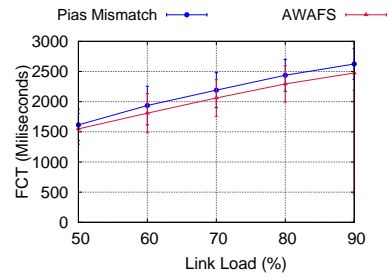}
          \caption{Tail FCT for Long Flows - Hadoop Workload}
          \label{fig:taillonghadoop}
      \end{figure}
      
  \end{minipage}
  \end{minipage}

\subsubsection{Cache Workload}
 
In this section, we present the results of the experiments considering the Cache workload. For these experiments, PIAS was configured with the thresholds derived for the Data Mining workload.

Figures \ref{fig:averageshortcache} and \ref{fig:tailshortcache} presents the result of evaluating the FCT for the short flows. AWAFS outperformed PIAS and reduced the average FCT in almost 11.8\% and the tail FCT in almost 12.7\%.

 \begin{minipage}{\linewidth}
  \centering
  \begin{minipage}{0.45\linewidth}
      \begin{figure}[H]
          \includegraphics[width=150pt]{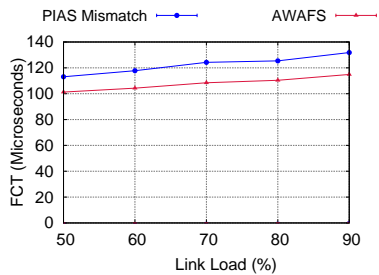}
          \caption{Average FCT of Short Flows - Cache Workload}
          \label{fig:averageshortcache}
      \end{figure}
  \end{minipage}
  \hspace{0.05\linewidth}
  \begin{minipage}{0.45\linewidth}
      \begin{figure}[H]
          \includegraphics[width=150pt]{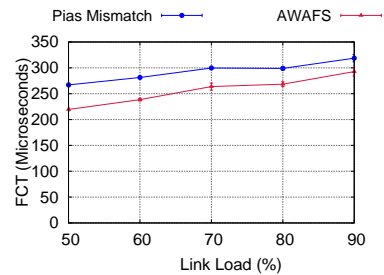}
          \caption{Tail FCT of Short Flows - Cache Workload}
          \label{fig:tailshortcache}
      \end{figure}
      
  \end{minipage}
  \end{minipage}
\vspace{0.5cm}

Figures \ref{fig:averagelongcache} and \ref{fig:taillongcache} present the results of assessing long flows. AWAFS improved both of the metrics, reducing the average FCT in 10.2\% and the tail FCT in 18.5\%.

 \begin{minipage}{\linewidth}
  \centering
  \begin{minipage}{0.45\linewidth}
      \begin{figure}[H]
          \includegraphics[width=150pt]{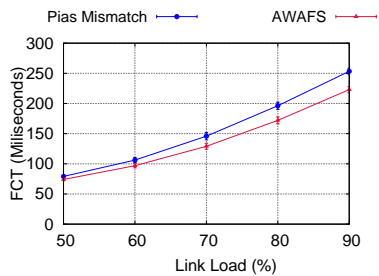}
          \caption{Average FCT of Long Flows - Cache Workload}
          \label{fig:averagelongcache}
      \end{figure}
  \end{minipage}
  \hspace{0.05\linewidth}
  \begin{minipage}{0.45\linewidth}
      \begin{figure}[H]
          \includegraphics[width=150pt]{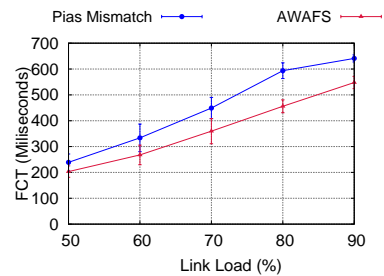}
          \caption{Tail FCT of Long Flows - Cache Workload}
          \label{fig:taillongcache}
      \end{figure}
      
  \end{minipage}
  \end{minipage}

\vspace{0.5cm}

Finally, Figures \ref{fig:averagecache} and \ref{fig:tcptimeoutscache} present the results for the overall average FCT and TCP timeout events respectively. AWAFS outperformed PIAS by reducing the overall average FCT around 5.3\% at high traffic load and the TCP timeouts count, which were actually reduced in almost 63\%.

 \begin{minipage}{\linewidth}
  \centering
  \begin{minipage}{0.45\linewidth}
      \begin{figure}[H]
          \includegraphics[width=150pt]{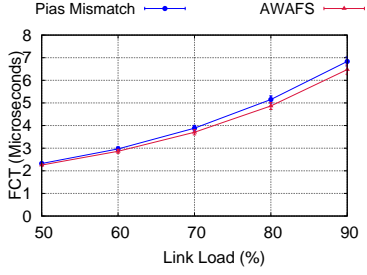}
          \caption{Overall Average FCT - Cache Workload}
          \label{fig:averagecache}
      \end{figure}
  \end{minipage}
  \hspace{0.05\linewidth}
  \begin{minipage}{0.45\linewidth}
      \begin{figure}[H]
          \includegraphics[width=150pt]{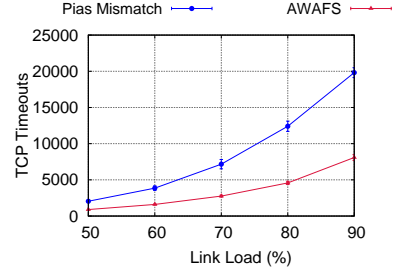}
          \caption{TCP timeouts Count - Cache Workload}
          \label{fig:tcptimeoutscache}
      \end{figure}
      
  \end{minipage}
  \end{minipage}

\subsection{AWAFS vs PIAS with Heterogeneous Traffic}
\label{heter}
In our previous simulations, the traffic patterns were homogeneous, given that all the nodes generated traffic following the same flow size distributions during the experiments. In this section, we present the results of comparing AWAFS against PIAS \cite{bai_pias:_2017} in a scenario with traffic heterogeneity.
\\
\\
In the original topology of 144 hosts, we have 144 x 143 communication pairs in total. In order to create the heterogeneous traffic pattern, in each link (i, j), we generated traffic according to the Web Search \cite{alizadeh_data_2010} workload if $i < j$. Otherwise, traffic was generated according to the Data Mining workload \cite{greenberg_vl2:_2009}. In this way, different links had different patterns of traffic. This situation clearly generated the scenario of threshold mismatch. In our experiment we compared PIAS with the thresholds derived for the Web Search workload (configured in all the links) and AWAFS with the parameters used in the previous experiments. Demotion thresholds for PIAS were generated following the same approach used in Section V-C number 4 of \cite{bai_pias:_2017}.
\\
\\
Figures \ref{fig:shortheter} and \ref{fig:tailshortheter} present the average and tail FCT respectively for short flows for different link occupation levels. It can be seen that AWAFS outperformed PIAS regardless the traffic load. It is important to remark the situation of threshold mismatch inevitably introduced by the traffic heterogeneity. In this case, the average FCT for short flows presented improvements between 4.3\% and 5.1\% whereas for the tail FCT the improvements were between 8\% and 15\%.
\\
\\

 \begin{minipage}{\linewidth}
  \centering
  \begin{minipage}{0.45\linewidth}
      \begin{figure}[H]
          \includegraphics[width=150pt]{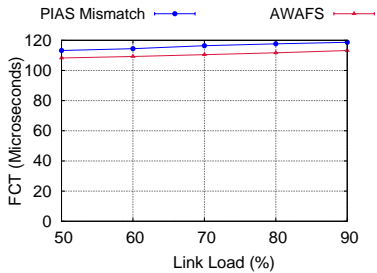}
          \caption{Average FCT for Short Flows - Herogeneous Traffic}
          \label{fig:shortheter}
      \end{figure}
  \end{minipage}
  \hspace{0.05\linewidth}
  \begin{minipage}{0.45\linewidth}
      \begin{figure}[H]
          \includegraphics[width=150pt]{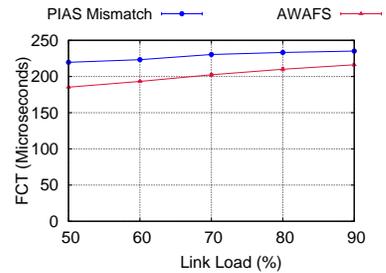}
          \caption{Tail FCT for Short Flows - Heterogeneous Traffic}
          \label{fig:tailshortheter}
      \end{figure}
      
  \end{minipage}
  \end{minipage}

\vspace{0.5cm} 
These results show the robustness of AWAFS, specially regarding its capacity to adapt to variations on the traffic behavior. Results presented in the previous sections showed how AWAFS can react to time variations in the traffic, caused by the presence of different workloads in different times. Results presented on this section showed that AWAFS can also react and adapt itself to space variations in the traffic patterns, which are originated by the presence of different traffic patterns in different zones of a network. Thus, our results are promising in the sense of introducing AWAFS as a mechanism to improve the performance of cloud applications in large scale data center networks.

\subsection{Summary}
\label{summary}
 
In this section, we have presented a comparison of AWAFS and its closest related work, PIAS. We have observed that in general, AWAFS introduces benefits for a variety of workloads by reducing both the average and tail FCT. This reduction is specially notorious for short flows, although it can benefit also medium and long flows. In those cases where these medium and long flows are not improved, the performance penalty is not significant for the overall performance since these flows are not a representative amount within the respective workloads. Also, in the case of long flows, the increment of the FCT is irrelevant considering the average FCT and the size of these flows in comparison with short flows (For example, long flows for the Data Mining workload might have sizes up to 600MB).

It is important to remark that AWAFS achieves these results leveraging its adaptability, based on traffic observation and its complete agnosticism since it does not require any a-priori information about the workloads.

\section{Discussion}
\label{discussion}

In the previous sections, we have assessed different aspects of the operation of AWAFS. We divided the evaluation in two main simulation scenarios consisting in a proof-of-concept and the operation in a Large Scale topology. In the first scenario, we observed how AWAFS effectively adjusts the demotion threshold converging towards the size of the short flows present in the workload. In this case, the convergence was exact due to the fact that the size of the short flows was fixed. 
This scenario allowed us to verify that the mechanism of monitoring based on calculating a set of percentiles on the list of completed flow sizes in a given time window provides an adequate hint to adapt the demotion thresholds. Through this calculation, we can adapt the demotion thresholds in a MLFQ scheduler. This adjustment provides the minimization of the FCT, specially when compared with the cases where there exists mismatch between the configured threshold and the current workload present in the network. We also could verify that the prioritization of the short flows does not cause starvation on the long flows as we did not observe increments on the metrics assessed for these flows. 

The Large Scale scenario allowed us to assess AWAFS in more realistic conditions. In this scenario, we considered the simulation of a larger topology with traffic based on workloads observed in production data center network environments. When the demotion thresholds can be somehow configured with optimal values, AWAFS does not introduce relevant overhead in the operation since it does not increment considerably the FCT of the short flows. On the other hand, when there is mismatch on the demotion thresholds, AWAFS improves the performance of the workload by minimizing the FCT. As we observed, AWAFS offers better results with an intermediate number of queues. With 4 queues, our approach introduces an important improvement in the FCT when compared to the static configuration. Larger number of queues still introduces improvement but this is less notorious.

By using percentiles as hint to adapt the demotion thresholds, AWAFS is one step closer to be a truly agnostic flow scheduler. Indeed, since it does not require specific a-priori information about the properties of the workloads in order to plan the scheduling, AWAFS is truly workload-agnostic. The rationale used to conceive the mechanism of threshold adjustment is consistent with the properties that have been identified and reported in the literature for the traffic of typical data center applications. Our experiments have shown that this approach to adjust the demotion thresholds is effective in terms of improving the FCT of short flows, and adapting autonomously to variations in the workload. These two aspects constitute a contribution to the state-of-the-art considering that we address two limitations of the closest related work. As we have previously described, AWAFS does not require any a-priori information about the workloads and it can react and adapt autonomously to changes in the traffic present in the network. 

Finally, the comparison between AWAFS and PIAS allowed us to confirm the effectiveness of the adaptability proposed in our approach. We could observe that even in a scenario where different traffic patterns coexisted in different zones of the network, AWAFS could adapt itself accordingly. With this adaptability, it was possible to outperform PIAS, since its static configuration was optimal for some switches but it was suboptimal for others. 

\section{FINAL REMARKS}
\label{Conclusions}
\subsection{Conclusions}

In this paper, we presented AWAFS, an Adaptable Workload-Agnostic Flow Scheduling mechanism, inspired by a state-of-the-art solution based on agnostic flow scheduling. Our approach is adaptable in the sense that its operation is based on observation of the traffic present in the network in order to autonomously adjust the configuration of the flow scheduling. Also, it is workload-agnostic in the sense that it does not require prior information about the workloads. Thus, AWAFS overcomes a limitation present on the state-of-the-art which is the requirement of information about the CDF of the workload that will be present in the network.

We evaluated AWAFS via simulation, both in a proof-of-concept in order to verify its operation, and in a large scale data center topology. In our experiments, we observed that AWAFS does minimize the FCT of short flows without inducing starvation on long flows. We also verified the adaption capability of AWAFS by observing how it adjusts its scheduling configuration when the workload in the network changes. Finally, we confirmed that AWAFS still provides minimization of the short flows at high traffic loads. Despite this is not a common situation on data center networks \cite{roy_inside_2015}, we observed that AWAFS improves the FCT in a wide range of traffic loads. 

We made evident that the combination of local (acquired at switches or packet transmission component at end hosts) and remote (at switches) information enables the adjustment of the flow scheduling component without requiring prior information about the workload properties. Hence, we made a step in the field of agnostic flow scheduling. Due to its adaptability, this scheme is promising to be used on general purpose data centers where multiple different workloads can be present. Also with the advent of hardware such as programmable forwarding devices \cite{bosshart_p4:_2014,sivaraman_programmable_2016} there is a wider landscape to evolve and improve this solution.

\subsection{Future Work}

Despite the positive results obtained, AWAFS still can not be considered as a complete solution. Hence, many aspects of the design of AWAFS are left as work to be developed in the future, in order to further improve its operation and adapt it to the new scenarios introduced in the state-of-the-art in data center networking.

One aspect to be considered for future work is to provide a smarter mechanism for the definition of the reference percentiles. In the current design of AWAFS, these percentiles are provided as parameters based on the understanding of the data center traffic properties. An important improvement that could be applied to our proposal is the capacity to determine from the observation of the traffic which would be adequate values to define the percentiles, in order to increase the accuracy of the scheduler. Proposals such as AuTO \cite{chen_2018} suggest a promising research line, considering the incorporation of Artificial Intelligence techniques to address the flow scheduling problem.

Another aspect to consider is the optimization of the size of the data structure used within the forwarding device to store the sizes of the completed flows. In our design, we assumed  such list can be available. There are important proposals in the literature introducing the notion of programmable hardware which could be leveraged for complex operations such as the implementation of this list \cite{sivaraman_packet_2016,sivaraman_programmable_2016}. 

Finally, an important task to consider as future work is the implementation of AWAFS in an actual forwarding device. Proposals available in literature \cite{sharma2018approximating,kaljic2019survey, pontarelli2019flowblaze,saeed2019eiffel,he2018exploring, castanheira2019flowstalker, kundel2018p4, scholz2019cryptographic, sivaraman2016packet} show that there exists elements within current forwarding devices which can be leveraged in order to have a more mature starting point to implement AWAFS.

\section*{Acknowledgment}
This work was developed with the support of Universidad Nacional de Colombia (UNAL) through the scholarship "Outstanding Postgraduate Student" during years 2012 to 2016, Colombian Ministry of Science and Technology (MinCiencias), through the scholarship "567 - National Doctorate Studies", during years 2013 to 2019, Universidad de Antioquia (UDEA) and Universidade Federal do Rio Grande do Sul (UFRGS), Porto Alegre, Brazil.

\bibliographystyle{unsrt}  
\bibliography{template}  

\end{document}